# SU(2) potentials in quantum gravity

W. Beirl, B.A. Berg[a] [*], B. Krishnan[†], H. Markum and J. Riedler

Institut für Kernphysik, Technische Universität Wien, A-1040 Vienna, Austria

[a]Department of Physics and SCRI, The Florida State University, Tallahassee, FL 32306, USA

We present investigations of the potential between static charges from a simulation of quantum gravity coupled to an SU(2) gauge field on $6^3 \times 4$ and $8^3 \times 4$ simplicial lattices. In the well-defined phase of the gravity sector where geometrical expectation values are stable, we study the correlations of Polyakov loops and extract the corresponding potentials between a source and sink separated by a distance $R$. In the confined phase, the potential has a linear form while in the deconfined phase, a screened Coulombic behavior is found. Our results indicate that quantum gravitational effects do not destroy confinement due to non-abelian gauge fields.

## 1. INTRODUCTION

The biggest problem in constructing a grand unified theory of all fundamental forces is to include the gravitational interactions because one encounters severe difficulties in quantizing gravity. The perturbative treatment suffers from the well-known problems of the unboundedness of the action and non-renormalizability. Regge Calculus provides a non-perturbative way for investigations of the Euclidean Einstein action, called Regge–Einstein action in this context, on a lattice and offers the possibility to construct a unified theory by coupling gauge fields to the skeleton. One of the interesting results of such investigations was the discovery of an "entropy dominated" smooth phase in quantum gravity, where the expectation values with respect to the pure Regge–Einstein action are stable [1]. The next logical question was to address the physical relevance of this regime. Efforts in this direction have been made by coupling a non-abelian gauge field to gravity [2]. If one assumes that the world without gravity is described by a grand unified asymptotically free theory, these numerical studies investigate the relation of the hadronic scale to the Planck scale. In particular, one is interested whether confinement exists and hadronic masses (deconfinement temperature, string tension, etc.) may be chosen small compared to the Planck mass. In this work, we perform a non-perturbative, finite temperature study of the behavior of the static quark potentials in the coupled system of quantum gravity and SU(2) gauge fields in four spacetime dimensions and extract a value for the string tension in the confined phase.

[*]Supported in part by DOE under Contracts DE-FG05-87ER40319 and DE-FC05-85ER2500.
[†]Lise-Meitner Postdoctoral Research Fellow sponsored by FWF under Project M078-PHY.

## 2. THE MODEL

In Regge Calculus the edge lengths are taken to be the dynamical variables of the discretized spacetime manifold and in four dimensions the curvature is concentrated on triangles. We choose a hypercubic triangulation with $N_s^3 \times N_t$ vertices and a scale-invariant measure in our simulations:

$$D[\{l^2\}] = \prod_l \frac{dl^2}{l^2}, \qquad (1)$$

where $l$ is used to denote the link label as well as the length. The system of SU(2) gauge fields coupled to quantum gravity has the action [2]

$$S = 2m_p^2 \sum_t A_t \alpha_t - \frac{\beta}{2} \sum_t W_t \, \text{Re}[\text{Tr}(1 - U_t)], \quad (2)$$

with the bare Planck mass $m_p$ and the gauge coupling $\beta$. $A_t$ and $\alpha_t$ correspond to the area and the deficit angle of the triangle $t$ while the weights $W_t$ describe the coupling of gravity to the gauge field. $U_t$ is the ordered product of SU(2) matrices around $t$.

The Polyakov loop $P(R)$ in the short extent $L_t$ of the lattice describes the propagation of a heavy quark and acts as an order parameter. As in conventional SU(2) lattice gauge theory at finite temperature $T$, we extract from the correlation function

$$\langle P(0)P^\dagger(R)\rangle = \exp[-\frac{1}{T}V(R)], \qquad (3)$$

the quantity $V(R)$ corresponding to the potential between the heavy quark-antiquark pair. In the confinement phase, $V(R)$ should grow linearly for large $R$ due to an infinite free energy of isolated quarks, while in the deconfinement phase, one expects a screened Coulombic behavior:

$$\begin{aligned} V(R) &= \frac{-\alpha}{R} + \sigma R \qquad \text{(confinement)} \\ &= \frac{-\alpha}{R}\exp(-\mu R) \quad \text{(deconfinement)}, (4) \end{aligned}$$

where $\sigma$ is the string tension, $\alpha$ the Coulomb parameter and $\mu$ the Debye screening mass.

We define a length scale by setting the expectation value of the volume of a pentahedron $\langle v_p\rangle$ to be a constant. The quantity $l_0 = \langle v_p\rangle^{\frac{1}{4}}$ serves as a length unit. In the well-defined phase of gravity expectation values are stable and the physical Planck mass $m_P$ can be related to $l_0$ by $m_P = m_p \times l_0^{-1}$. In the context of quantum gravity, there exist reasons to consider a fundamental length [3] as physical, which then defines a natural cutoff. For the hadronic masses $m_h$, one expects an asymptotic scaling law from the renormalization group equation [2]. If this scaling could be observed, our simulations relate to the physical region in the same way as standard lattice gauge calculations do. In addition, classical gravity should be recovered for macroscopic distances.

To address the issue of the two different length scales, the Planck mass should turn out large in comparison to the hadronic mass. In our finite temperature $L_s^3 \times L_t$ lattices, decreasing the length unit $l_0$ means increasing the gauge coupling which is restricted by a critical value, $\beta < \beta_c$. To allow larger $\beta_c$ values the temporal extension has to be enlarged, making sure simultaneously that the smooth phase of gravity does not break down, $m_p^2 < m_c^2$, with the border of the smooth phase $m_c^2 > 0$. In this study, we extend the previous $N_t = 2$ investigation [2] to $N_t = 4$. We find that $m_c^2 \approx 0.025$ is almost unaffected by the accompanying increase in $\beta_c$. The corresponding deconfining temperature is defined by $T_c = 1/(N_t\langle x_l\rangle_t)$ with $\langle x_l\rangle_t$ the link length in the short direction of the skeleton. This gives from $N_t = 2$ to $N_t = 4$, an increase in the mass ratio $m_c/T_c$ by about a factor two.

## 3. NUMERICAL RESULTS

We performed our simulations on a $6^3 \times 4$ lattice first and then, to extract a value of the string tension, on an $8^3 \times 4$ lattice. We measure the potentials between external sources by computing the correlations of Polyakov loops along the main axes. Fig. 1(a) presents the static quark potentials in the presence of gravity, whereas Fig. 1(b) depicts the situation with gravity switched off. To the best of our knowledge, Fig. 1(b) is the first investigation of SU(2) potentials on a flat simplicial lattice. Fig. 1(a) shows that $V(R)$ with $m_p^2 = 0.005$ in the "entropy dominated" phase behaves very differently in the two phases of the gauge field. For $\beta = 1.55$ the potential rises steeply with $R$, while for $\beta = 1.60$ it is relatively flat suggestive of a screened Coulombic behavior. In Fig. 1(b) we compare with the potentials from pure gauge simulations on the flat simplicial lattice at two values of $\beta$ below and above the deconfinement transition ($\beta = 1.02$ and $\beta = 1.10$). Qualitatively, the shape of the potentials is similar to that of Fig. 1(a). The solid lines are fits to the potential according to Eq. (4). The fit parameters of the deconfined phase will be discussed elsewhere. On the fluctuating lattice, a value of the string tension $\sigma_{grav}^{simp} = 0.454(9)\langle x_l\rangle_s^{-1}\langle x_l\rangle_t^{-1}$ is extracted from the first three points with $\alpha = 0$. The average link along the main axes has a value of $\langle x_l\rangle_s \approx \langle x_l\rangle_t \approx 3.3 l_0$. Our data in the confined phase resolve practically no Coulomb effects in the potential between the quark sources. It is too early to comment on whether quantum gravitational effects may be responsible for such a behavior. For the pure gauge case, a string tension value of $\sigma_{flat}^{simp} = 0.389(7)a^{-2}$ is obtained with $\alpha = \pi/12$. The lattice spacing $a$ on the flat lat-tice can also be expressed in units of $l_0$, $a \approx 2.2 l_0$.

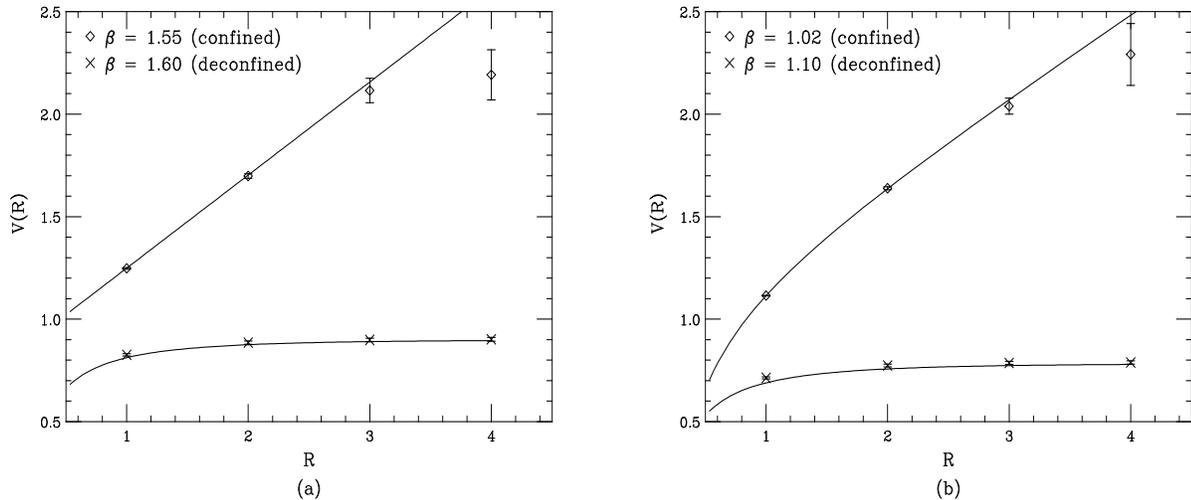

Figure 1: Static quark potentials on an $8^3 \times 4$ fluctuating lattice with $m_p^2 = 0.005$ (a) and for a flat simplicial lattice with gravity switched off (b).

Finally, we comment on the distance $R$ between quark sources on a fluctuating skeleton. The correct distance between two points should be measured using geodesic distances. The geometry is known to be Euclidean only inside the pentahedron with the curvature on the triangles. We take the distance $R$ between the source and sink to be equal to the index distance along the main axes of the skeleton. This seems a reasonable approximation in the well-defined phase with small curvature fluctuations. Other proposals using the methods of scalar field propagation [4] require corrections at small distances and turned out to be unsuitable for our purposes.

## 4. CONCLUSION

We have studied static SU(2) quark potentials in the presence of quantum gravity and find that confinement from the non-abelian gauge fields is not destroyed by quantum gravitational effects. In the confined phase we get a potential linearly rising with $R$, and in the deconfined phase we see a behavior resembling a screened Coulomb potential. We have extracted string tension values in the confined phase for both the coupled system of quantum gravity and SU(2) gauge fields and for the pure gauge theory on a simplicial lattice without gravity. We have noted that the "entropy dominated" phase seems to be stable with the increase in $N_t$. A decrease in the ratio $T_c/m_c$ by a factor two was observed. To present evidence for a further fall-off in this ratio would require simulations on lattices with larger $N_t$ extensions.